**An Open-Source Web-Based Viewer for Disseminating Image-Derived Models**

Christian Herz[*,a], Kyle Sunderland[*,b], Matthew Daemer[a], Julia Iacovella[a], Sehdev Pressley[a], Ashley Koluda[a], Csaba Pinter[c], Kevin Zuber[d], John Daniels[d], Allen Ladd[d], Mali Hetmaniuk[a], Kevin K. Whitehead[e], Andras Lasso[b], Matthew A. Jolley[a,e]

*These authors contributed equally to this work.

**Affiliations**

a Department of Anesthesiology and Critical Care Medicine, Children's Hospital of Philadelphia, 3401 Civic Center Blvd, Philadelphia, PA, USA.

b Laboratory for Percutaneous Surgery (PerkLab), School of Computing, Queen's University, Kingston, ON, Canada.

c Ebatinca S.L., Las Palmas de Gran Canaria, Spain.

d Research Information Services, Children's Hospital of Philadelphia, 3401 Civic Center Blvd, Philadelphia, PA, USA.

e Division of Cardiology, Children's Hospital of Philadelphia, 3401 Civic Center Blvd, Philadelphia, PA, USA.

**Word Count:** 4,050

**Corresponding Author**

Matthew A. Jolley, MD

Department of Anesthesiology and Critical Care Medicine and Division of Cardiology

Children's Hospital of Philadelphia

Associate Professor, University of Pennsylvania

3401 Civic Center Blvd

Philadelphia, PA 19104, USA

Phone: 215-280-5643

Email: jolleym@chop.edu

## Acknowledgements

We thank Research Information Services (RIS) at Children's Hospital of Philadelphia (CHOP) for providing the computing infrastructure and technical support that made this work possible. We also acknowledge the Research Institute at Children's Hospital of Philadelphia for its institutional support of this research.

## Statements and Declarations

**Funding:** This work was supported by the National Institutes of Health (NIH) grant 1R01HL153166, a Children's Hospital of Philadelphia Post-Frontier Award, the Cora Topolewski Fund at the Children's Hospital of Philadelphia Pediatric Valve Center, the Topolewski Endowed Chair in Pediatric Cardiology, and The Mussallem Congenital Heart Disease Alliance.

**Competing Interests:** Csaba Pinter is an employee of Ebatinca, an open-source based software company. The other authors have no relevant financial or non-financial interests to disclose.

**Author Contributions:** All authors contributed to the study conception and design. Software design and implementation were performed by Christian Herz, Kyle Sunderland, Andras Lasso, and Csaba Pinter. Enterprise deployment, virtualization infrastructure, and authentication and security configuration were performed by Allen Ladd, Kevin Zuber, and John Daniels. Patient-specific model creation, scene preparation, and development of the clinical modeling workflow were performed by Matthew Daemer, Julia Iacovella, Sehdev Pressley, and Ashley Koluda. Clinical requirements definition, end-user evaluation, and validation were performed by Christian Herz, Kyle Sunderland, Andras Lasso, Csaba Pinter, Matthew Daemer, Kevin K. Whitehead, and Matthew A. Jolley. Mali Hetmaniuk contributed to the development of educational applications of the Viewer. The study was supervised by Andras Lasso and Matthew A. Jolley. The first draft of the manuscript was written by Matthew A. Jolley, Christian Herz, Kyle Sunderland, Matthew Daemer, and Julia Iacovella, and all authors commented on previous versions of the manuscript. All authors read and approved the final manuscript.

**Ethics Approval:** This study was determined to be exempt by the Children's Hospital of Philadelphia Institutional Review Board, as a technical description using de-identified data, and was performed in line with the principles of the Declaration of Helsinki.

**Consent to Participate:** Informed consent requirement was waived per IRB.

**Consent to Publish:** Not applicable. No identifiable images are included.

**Data and Code Availability:** The Viewer module is open source and freely available as part of the SlicerHeart extension, which can be installed through the 3D Slicer Extension Manager. Its source code is available at https://github.com/SlicerHeart/SlicerHeart.

# An Open-Source Web-Based Viewer for Disseminating Image-Derived Models

Article Type: Technical Note

**Abbreviations:**

2D = 2-dimensional

3D = 3-dimensional

3DE = 3-dimensional echocardiography

4D = 4-dimensional (3D + time)

CHD = congenital heart disease

MRI = magnetic resonance imaging

CT = computed tomography

DICOM = Digital Imaging and Communications in Medicine

PHI = protected health information

RDP = Remote Desktop Protocol

## Abstract

Patient-specific 3-dimensional (3D) models are increasingly central to planning surgery and catheter-based intervention in children with congenital heart disease (CHD), whose anatomy is often atypical and variable. These models inform care only when the surgeons and interventional cardiologists who make treatment decisions can rotate, crop, and measure them directly. Desktop modeling applications require local installation, capable hardware, and training. Commercial web-based viewers are portable but closed, and, in our experience, cannot display volume rendering, time-resolved (e.g., 4-dimensional) data, or the custom workflows centers build for CHD. Because CHD populations are small and heterogeneous, such workflows are unlikely to be prioritized commercially, and end users have no means to add them. To address these limitations, we developed the SlicerHeart Viewer, an open-source module that presents a simplified interface to a complete 3D Slicer/SlicerHeart session running on a server. Clinicians access models created by clinical modeling services from any device through a web browser. Because 3D Slicer runs underneath, volume rendering, time-resolved data, and any SlicerHeart workflow, including those developed for CHD, are available without modification. The module is independent of the hosting environment; we chose internal servers behind institutional single sign-on, and a cloud-hosted option is in development. Over one year at a single pediatric center, nearly all of 82 modeled cases were shared through the Viewer with clinicians from five specialties, with cases loading in 5 to 10 seconds. The approach is not specific to CHD and is deployable at other centers.

## Background

Although congenital heart disease (CHD) is a common birth defect, it comprises small and heterogeneous lesion groups in which the anatomy of an individual patient is frequently atypical [1, 2]. Surgical and catheter-based treatment is correspondingly individualized, designed around that patient's arrangement of chambers, valves, and great vessels. Planning such interventions increasingly relies on multimodality 3-dimensional (3D) imaging and on the patient-specific 3D models derived from it, which help the heart team appreciate spatial relationships, practice a repair, and size devices before a procedure [3, 4]. Because cardiac anatomy is unique to the patient, a patient-specific model conveys understanding that cannot be drawn from a textbook or a prior case.

Models are typically generated from clinical images by segmentation of the anatomy of interest, work performed in commercial applications [5] or in any of a growing number of open-source platforms, including 3D Slicer [6], ITK-SNAP [7], the Medical Imaging Interaction Toolkit [8] and SimVascular [9]. Many centers use several in combination, so the models reaching a clinician are heterogeneous in format and origin. Among these, 3D Slicer has been extended for cardiac applications by SlicerHeart [3], an open-source extension supporting import and visualization of 3-dimensional echocardiography (3DE), volume rendering, valve segmentation, device placement, and modeling workflows developed for complex congenital anatomy, including virtual planning of the intracardiac baffle [10, 11]. Because SlicerHeart operates on the 3D Slicer scene, models produced in other applications can be imported and reviewed alongside those created within it, so a case assembled from several sources can be viewed together.

Creating the model, however, delivers value only when the surgeons and interventional cardiologists who make treatment decisions can explore it in a form that informs patient care. Physical 3D prints add cost and are time consuming, and static images or videos are not interactive. The full 3D Slicer environment is a desktop application requiring local installation, hardware, and training, which is impractical for a clinician reviewing a case on a phone or at the bedside. Commercial web-based viewers make patient-specific models more portable and are practical in routine use; however, in our experience, they lack capabilities important for congenital cases, including volume rendering, support for time-resolved (4-dimensional (4D)/5-dimensional (5D)) data, and the ability to allow the end user to create novel custom workflows [12]. Additionally, institutions often prefer that patient-specific models be hosted within their own security infrastructure rather than on an externally managed commercial platform.

In order to address these limitations, we developed the SlicerHeart Viewer: a lightweight, browser-based viewer that allows clinicians to interactively review patient-specific cardiac models on a desktop computer, tablet, or phone without local installation. It presents a simplified interface to a complete 3D Slicer/SlicerHeart session, preserving its full feature set, including volume rendering, time-resolved data, and custom workflows. To satisfy institutional security requirements, we deployed the Viewer on an enterprise back end (server) behind our institution's authentication service, enabling controlled review of patient-specific models by the care team. We describe the design and implementation of the Viewer and report our initial experience deploying it within our institution's cardiac clinical modeling program.

## Methods

### Software platform and implementation

The SlicerHeart Viewer is implemented as a scripted module within SlicerHeart [3], the open-source extension of the 3D Slicer platform [6] for image-based cardiac modeling, analysis, quantification, and procedure planning. Like other SlicerHeart modules, the Viewer is written in Python and builds on the same underlying libraries, including the Visualization Toolkit (VTK) for visualization, the Insight Toolkit (ITK) [13] for image storage and processing, the Qt toolkit for graphical user interface, and Medical Reality Modeling Language (MRML) for data representation [6]. Segmentations, surface models, volume renderings, markups, transforms, and image sequences produced by any SlicerHeart workflow are therefore available to the Viewer without conversion or export. The Viewer is not a reimplementation of a 3D display; it is a simplified interface presented on top of a complete 3D Slicer session running in the background. Consequently, any capability available in 3D Slicer or SlicerHeart can be surfaced through the Viewer, and additional capabilities can be added as needed without altering the delivery architecture. Like all other modules in the SlicerHeart extension, the Viewer is free and open source.

### Design approach: a simple interface with access to the full platform

A central design goal was to reconcile two competing requirements: the interface had to be simple enough for clinicians to use without training, yet flexible enough to serve advanced users. By default, the Viewer presents a deliberately minimal, task-oriented interface that exposes only the controls needed to review a prepared case. At any time, however, users are free to switch from the Viewer to the full 3D Slicer interface.

### Preserved visualization and analysis capabilities

Because the Viewer has access to the full feature set of 3D Slicer and the SlicerHeart extension, all capabilities that are important for congenital cardiac cases are preserved. These include interactive volume rendering; visualization of time-resolved (4D and 5D) data; capped cross-sections; and presentation of models produced by SlicerHeart custom workflows, such as the baffle planner [10], transcatheter edge-to-edge repair (TEER) planning [14], and transcatheter device modeling [15]. End users can also perform interactive cropping and simple measurements directly within the Viewer, including linear length, curved length, and area.

### Scene views

To balance simplicity and flexibility, the Viewer relies heavily on the scene views feature of 3D Slicer. Scene views are snapshots of the loaded scene that a modeling technician prepares in advance and that the clinician can then step through with a single click from a list. Each scene view stores the view layout, the position, orientation, and region of interest of each view, and the visualization settings (e.g., visibility, transparency, color, and cropping) of each data object, such as a segmentation, a volume rendering, or a device positioned within image-derived anatomy. While the modeling technician pre-populates the scene views required for a given case, the clinician retains interactive control and can modify parameters such as cropping and take simple measurements. Scene views were introduced in direct response to end-user feedback because stepping through a curated sequence of prepared views is preferable to opening multiple simultaneous 3D Slicer application instances, given the increased computational load and hardware requirements.

### Remote delivery architecture

The Viewer runs 3D Slicer, with the SlicerHeart extension preinstalled, on a server and displays its interface in a standard web browser. Clinicians can therefore review cases on a desktop computer, tablet, or phone without installing any software (Fig. 1). This application-virtualization approach preserves the complete desktop feature set without reimplementing it for the web.

The initial implementation uses Omnissa Horizon [16], an application- and desktop-virtualization platform for which our institution holds a license and has established operational familiarity. The server runs Microsoft Windows Server 2025 Datacenter and is equipped with 64 central processing unit (CPU) cores, 4 NVIDIA L40S graphics processing units (GPUs), 768 GB of random-access memory (RAM), and 4 TB of solid-state storage, hosted through Omnissa Horizon as a cluster of 8 virtual machines.

During the clinical modeling workflow, a web uniform resource locator (URL) is generated to provide browser access to the Viewer. The link encodes an application identifier (applicationId=Clinical+Viewer) that identifies the published application and prompts Omnissa Horizon to route the request to an available virtual machine, where a Windows batch script is executed. The script evaluates the arguments passed in the link and initiates the corresponding behavior; in the current configuration, the link specifies the file path of a stored 3D Slicer scene, and the batch script launches 3D Slicer, loads the specified scene, and displays the Viewer module. As a security measure, Omnissa Horizon exposes only the application's user interface without providing full desktop access.

The delivery layer is decoupled from the Viewer module. Because the module is an ordinary SlicerHeart component, the same interface can be served through alternative back ends. A purpose-built delivery layer supporting local-server and cloud deployment is in development, and we anticipate that native web-rendering frameworks, such as vtk.js-based approaches [17], trame [18], and emerging web-based frameworks will provide additional options as they mature.

**Authentication and secure access**

Access to the Viewer is mediated by our institutional authentication service, which uses the Lightweight Directory Access Protocol (LDAP). Because the application is published through the enterprise virtualization environment and authenticated by institutional single sign-on, clinicians can log in securely from any supported device using their standard institutional credentials, without needing to establish a virtual private network (VPN) connection. This arrangement lowers the barrier to review at the point of care while keeping patient-specific models within the institution's own security infrastructure.

**Clinical modeling and preparation workflow**

Patient-specific models are created from clinical images using established SlicerHeart [3] and Materialise Mimics [5] workflows. When needed, models generated in Mimics are imported into the 3D Slicer scene, so that models from either environment can be reviewed together in the Viewer. For dissemination, a modeling technician prepares the relevant scenes for each case using a consistent 3D Slicer version. The prepared scene file is then uploaded to the main virtual machine of the cluster, which serves as the source for both data distribution and Viewer updates. The technician connects to a virtual-desktop session on this machine through the Remote Desktop Protocol (RDP) and uses the ViewerLinkCreator module in SlicerHeart to generate a shareable link to the locally stored scene. The link can then be shared via email or during clinical case conferences. Scene files are distributed to all virtual machines in the cluster every 15 minutes, and Viewer software updates applied to the main virtual machine are propagated to the remaining machines by a daily job. Cases are de-identified using REDCap [19, 20] record identifiers, and viewing of de-identified data acquired for clinical purposes was approved by the institutional review board (IRB).

### Software availability

The Viewer module and ViewerLinkCreator are available on GitHub as modules within the SlicerHeart extension for 3D Slicer [3].

## Results

### Deployment and use

Over a year of clinical use beginning in July 2025, 82 clinical cases were modeled by our institution's cardiac clinical modeling program, and a Viewer link was generated for nearly all of them, establishing the Viewer as the routine means of delivering patient-specific models to the care team. Cases were reviewed by clinicians from cardiac surgery, cardiology, interventional cardiology, cardiac anesthesiology, and echocardiography. Source imaging included computed tomography (CT), cardiac magnetic resonance imaging (MRI), and 3DE, some of which were time-resolved acquisitions. In several cases, models derived from different modalities were combined within a single scene. The Viewer was also used to present patient-specific models at institutional cardiac surgical conferences and multidisciplinary valve reviews; an overview of representative cases is provided in Online Resource 1. Clinicians accessed the Viewer from desktop computers, tablets, and phones; interactive review on a tablet is shown in Online Resource 2.

### Workflow and performance

Each case was typically prepared with 2–4 scene views. Once models were complete, setting up the scene views took approximately 5–10 minutes, and generating the Viewer link took seconds, after which the link was emailed to clinicians for review on any device at their convenience. In our experience, this compares with substantially longer preparation time for slide presentations with screenshots and videos of a model, and days to weeks for a physical 3D print to pass through print preparation, printing, and post-processing. After a link was opened, the case loaded in the browser in approximately 5–10 seconds.

**Viewer interface and interaction**

The default layout presented a 3D view alongside 3 synchronized multiplanar reformats of the source images, with a compact toolbar for navigation, measurement, and display settings (Fig. 2). For time-resolved data, playback controls allowed the user to play or step through the cardiac cycle, with the 3D view and reformats updating together.

Technician-prepared scene views were listed in a side panel and selected with a single click. In the case shown in Fig. 3a, 2 scene views presented the same 4D CT dataset as a segmented aortic valve model and as a volume rendering (Figs. 2a and 4b), so the clinician could move between representations without adjusting any settings. Within each scene view, clinicians retained interactive control. The display could be cropped with a region-of-interest box or clipping plane, with the cropped region reflected in the reformats (Fig. 3b), and linear, curved-length, and area measurements could be taken directly in the scene.

**Time-resolved and volume-rendered imaging**

The Viewer displayed time-resolved data that, to our knowledge, commercial web-based viewers could not display without conversion or export. For a patient with truncal valve regurgitation, 4D flow MRI was displayed as 5D data (3 spatial dimensions, time, and velocity), with flow streamlines rendered over a volume rendering of the anatomy. Clinicians could play the 18-phase acquisition to follow the regurgitant jet across the cardiac cycle and relate it to valve anatomy in the synchronized reformats (Fig. 4a). Volume-rendered 4D cardiac CT was reviewed in the same way to examine an aortic valve coaptation defect (Fig. 4b).

**Virtual device and procedure planning**

Virtual device placement was used to assess device fit, in both size and spatial relationship to surrounding structures, and to compare candidate sizes and models, plan positioning, and evaluate potential compression or obstruction of adjacent structures. For transcatheter procedures, it also informed the delivery route, the angle of approach, and whether a transcatheter or surgical approach was preferable. Because device models were reviewed within the full scene rather than as exported meshes, clinicians could rotate and crop the anatomy and check device position against the source images in the synchronized reformats. Transcatheter planning examples include a stent in a patient with superior sinus venosus atrial septal defect and partial anomalous pulmonary venous return (Fig. 5a) and a TriClip device (Abbott Medical, Abbott Park, IL) in a patient being evaluated for tricuspid transcatheter edge-to-edge repair (Fig. 5b). In surgical planning, for example, the fit of an Abbott HeartMate 3 left ventricular assist device (LVAD) was evaluated against the chest wall and lung as well as the heart, because the model included the rib cage and left lung (Fig. 6a). The same review extended to surgical reconstruction. In a patient being considered for biventricular repair, a virtual intracardiac baffle was created with the SlicerHeart baffle planner and displayed with a volume rendering of the cardiac MRI, giving the surgeon an approximate patch size as a starting point for intraoperative sizing (Fig. 6b).

### Educational use

Scene views were also prepared to demonstrate congenital septal defect anatomy. In cardiac MRI-derived models, the locations of ventricular septal defect (VSD) and atrial septal defect (ASD) subtypes were annotated together with adjacent landmarks, including the valve annuli, atrioventricular node, coronary sinus, and inferior vena cava (Fig. 7). Capped cross-sections displayed the cut myocardium as solid, preserving wall thickness and the spatial relationship of each defect to surrounding structures.

### User feedback and iterative refinement

Informal feedback from clinicians was positive, and their requests for additional functionality guided iterative development. Most notably, scene views were introduced in response to user feedback; presenting a curated sequence of prepared views within a single session was preferred over opening multiple simultaneous application instances, which would have increased computational load and hardware requirements.

## Discussion

We present the SlicerHeart Viewer, an open-source module in the SlicerHeart extension for 3D Slicer that allows clinicians to interactively review patient-specific cardiac models on a desktop computer, tablet, or phone through a standard web browser, without local installation, licensing, or training. Rather than reimplementing a 3D display for the web, the Viewer presents a simplified interface to a complete 3D Slicer/SlicerHeart session. The capabilities most relevant to congenital cardiac cases, including volume rendering, time-resolved (4D and 5D) display, capped cross-sections during cropping, and center-specific custom workflows, are therefore preserved rather than approximated or reinvented.

The rationale for this work follows from the growth of patient-specific modeling as a clinical service. What began at many centers as an ad hoc research activity is now an essential, request-driven workflow embedded in surgical and catheterization planning [21]. That growth has shifted the operational bottleneck. Workflows for model creation are now well established; however, rapidly delivering the content to the clinicians caring for a patient, in a clinically useful format, remains a practical challenge. Modeling expertise is often concentrated in a small team with specialized software and hardware, while the clinicians who need the result are distributed across operating rooms, catheterization laboratories, conference rooms, and their homes while on call. Any solution to this final delivery step must therefore assume a reviewer who has no specialized software or training and may be using limited hardware, such as a tablet or phone.

Existing dissemination approaches address parts of this problem but not all of it (Table 1). Physical 3D prints are valuable for rehearsal but add cost and delay and represent a single fixed configuration of the anatomy. Static images and videos are trivially shareable, yet they discard interactivity, and the view that matters is often not the view that was captured. The desktop 3D Slicer/SlicerHeart application preserves every capability but requires installation, computing resources, and training, and is unavailable on tablets or phones. Generic browser-based mesh viewers display exported surface geometry on any device, but in doing so they separate the model from the images it was derived from, discarding volume rendering, temporal information, and analytic context.

Commercial web-based viewers are the closest prior alternative, and, in our experience, were practical and reliable for routine review of surface models. Their limitations were of two kinds. The first was functional. As closed products, they could not be extended to desired use cases such as displaying volume-rendered image data, time-resolved 4D and 5D sequences, or capped cross-sections when a clipping plane is passed through a segmentation. Most consequentially, they could not present the center-specific workflows tailored to the needs of CHD [10, 14, 15]. Because such platforms are built around larger and more uniform markets, capabilities specific to CHD are unlikely to be prioritized, and the end user has no mechanism to add them [3].

The second limitation was one of governance. Commercial viewers are typically delivered as externally hosted services, so patient-specific models and their identifiers (if not anonymized) leave the institutional environment for infrastructure the institution does not control or audit, and the aggregate protected health information (PHI) held externally potentially grows with the scale and duration of the service. Our institution preferred that images and models remain within its own security infrastructure, even when de-identified. In addition, reliance on an external product creates a continuity risk, because if the product becomes unavailable the dissemination pathway and data archival disappear with it. An open-source tool hosted within the institution and authenticated through enterprise single sign-on addresses both concerns.

Because the Viewer operates on the native 3D Slicer scene rather than on an exported subset of it, it is agnostic to how the content was produced. Segmentations, surface models, volume renderings, markups, transforms, and image sequences derived from 3DE, CT, cardiac MRI, or rotational angiography can be presented together, as can models generated in commercial applications and imported into the scene. Any current or future SlicerHeart module can likewise be exposed without modifying the Viewer or the delivery layer, so that tools a center develops for its own population are immediately shareable with its own clinicians. This extensibility is what closed platforms cannot offer, and it matters most for the small, heterogeneous populations least likely to attract commercial development [3]. Scene views make that breadth usable, allowing a technician to curate what a clinician sees for a given case. Nothing in the design is specific to CHD, and the same architecture should serve adult structural heart disease and other specialties that rely on image-derived models.

The same infrastructure also lends itself to education and to engagement with patients and families. Because scenes are prepared once and opened by link on any device, a curated library of de-identified cases can accumulate alongside the clinical service, holding representative examples of common lesions together with the repairs and device positions used to address them. Such a library offers trainees, nurses, and sonographers an interactive alternative to textbook diagrams, reviewable on a phone during rounds or projected in a teaching conference without dedicated hardware. For families, the more compelling use is the patient-specific model itself. Explaining an unusual arrangement of the great vessels, or the plan for a complex intracardiac baffle, is difficult with 2-dimensional (2D) drawings, whereas a model of the child's own heart that a parent can rotate on a tablet conveys both the anatomy and the proposed repair directly. At present, these models are shown to families by their provider during counseling, so access remains within the existing clinical relationship. Extending access to families directly would require additional identity management, consent, and access-control infrastructure, which we regard as future work. We have not formally evaluated comprehension or satisfaction in either setting.

The Viewer module is also independent of the mechanism that delivers it. Omnissa Horizon was a pragmatic starting point rather than an architectural commitment, chosen because our institution held a license and could reach clinical use quickly. We are now developing a purpose-built, open-source delivery layer with two intended deployment models. The first is local-server deployment, in which the whole pathway runs inside the institutional network, minimizing latency and keeping images and identifiers within the existing security perimeter. The second is cloud deployment with secure hosting, removing the requirement that an adopting center maintain its own servers. Here we are particularly interested in a federated approach, in which PHI is removed before data leaves the originating institution and re-associated locally at the point of review, so that the hosted component holds only de-identified content. Such an architecture could allow a center with an established modeling program to deliver models to clinicians at remote sites that have none. These deployment models remain in development and have not been evaluated.

Several limitations should be stated. 3D Slicer, SlicerHeart, and the Viewer have not received regulatory approval. 3D Slicer is distributed explicitly as research software, and responsibility for compliance with local rules rests with the user. In this deployment, the Viewer is an adjunct to, and not a substitute for, review of source images within cleared clinical systems. We also note that the conditions making center-developed tools necessary in CHD, namely small heterogeneous populations and limited commercial investment, are the same conditions that make such tools hard to fit within frameworks built around uniform, high-volume adult-focused products.

Other limitations are technical and operational. Because rendering occurs server-side, usability depends on network quality, and performance over a constrained mobile connection is not equivalent to performance on the institutional network. Concurrent sessions consume server GPU and memory resources, and we have not characterized scaling behavior under load. The delivery layer described here currently depends on a commercially licensed virtualization platform, a dependency that the open-source back end is intended to remove. Scene preparation continues to require a trained modeling technician, so the Viewer lowers the barrier for the reviewer but not for model creation. Finally, this is a single-center experience over one year, evaluated through informal user feedback rather than a structured instrument, without measurement of effects on decision making or outcomes.

Several directions follow from this experience. The most immediate is closer integration with clinical imaging infrastructure, including saving models and scenes to Digital Imaging and Communications in Medicine (DICOM) and integrating the DICOM browser so that a case can be shared by patient identifier, with the relevant studies and series retrieved directly rather than by pointing to one prepared scene file. We also intend to support synchronized multi-user review, to capture Viewer measurements as structured data returned to the modeling report, and to pursue multicenter deployment with formal usability and clinical-impact evaluation. Native web-rendering paths such as vtk.js and trame may in time permit a hybrid approach in which lighter content is rendered on the client, and there are emerging options for running the full Slicer application in the web browser itself.

## Conclusion

We developed and deployed an open-source SlicerHeart module that brings full-featured, interactive patient-specific cardiac models to clinicians on any device through a web browser, lowering the barrier between the modeling team and the point of decision. Because it is open source, institutionally hosted, and preserves the complete SlicerHeart feature set, it offers a readily deployable path for other centers to disseminate patient-specific models in CHD and beyond.

## Figures

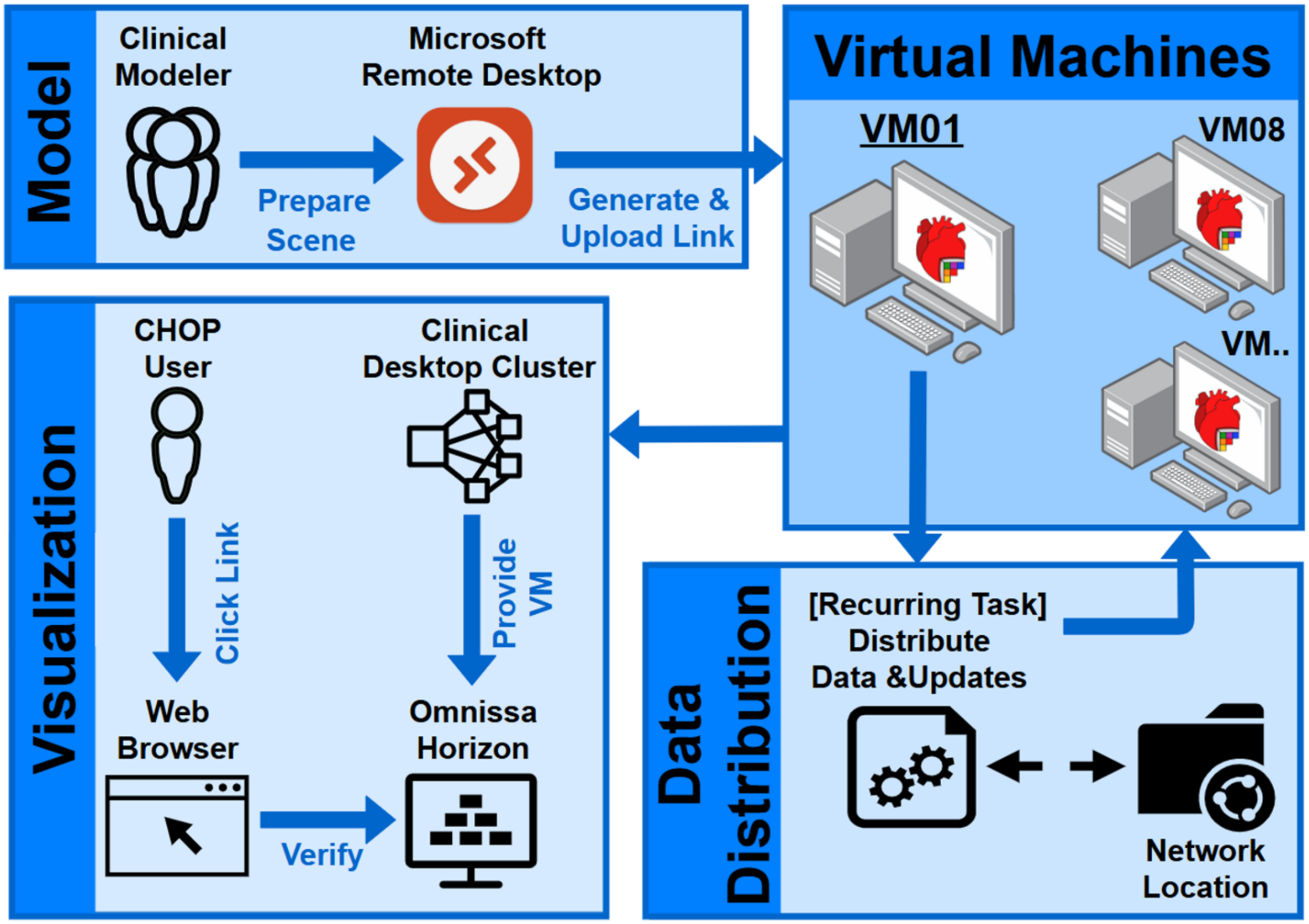

**Fig. 1** Current SlicerHeart Viewer deployment architecture and clinical workflow. Clinical modelers prepare case-specific scenes in 3D Slicer; the ViewerLinkCreator module generates shareable web links. Clinicians access the Viewer through a standard web browser connected via Omnissa Horizon virtualization, which authenticates users through institutional LDAP and routes requests to a Clinical Desktop Cluster. The cluster comprises GPU-equipped virtual machines that enable parallel case access. Automated synchronization ensures all users access current models and updates. This architecture allows clinicians to review patient-specific cardiac models across any device while maintaining institutional data security. GPU=graphics processing unit; LDAP=Lightweight Directory Access Protocol

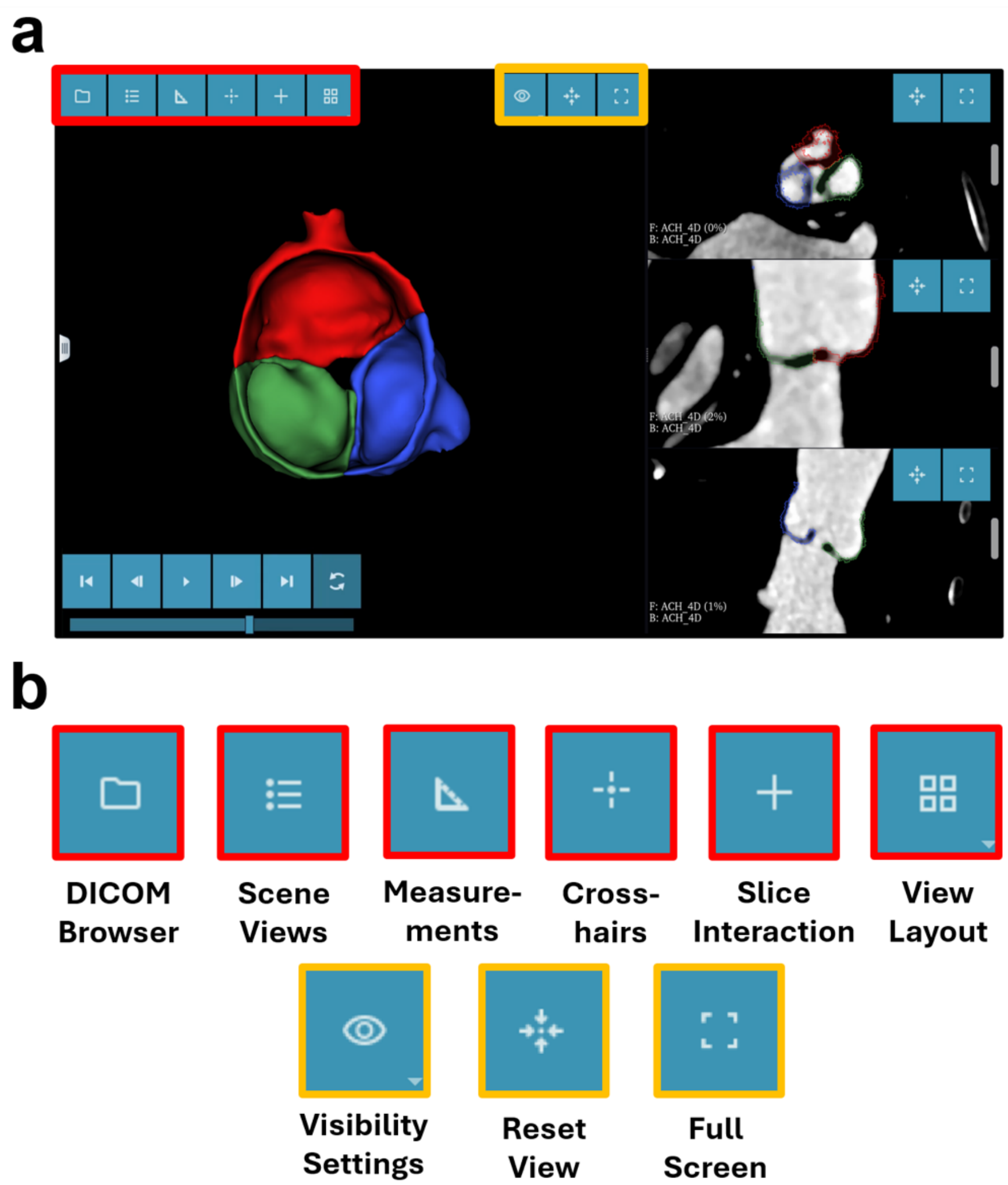

**Fig. 2** SlicerHeart Viewer interface for interactive review of patient-specific cardiac visualizations. (a) Left panel displays CT-derived aortic valve model; right panel shows synchronized multiplanar 2D reformats. Interactive manipulation tools are highlighted in red; visibility and rendering controls are highlighted in yellow. (b) Complete toolbar legend identifying viewer controls (red) and visualization options (yellow). 2D=2-dimensional; CT=computed tomography

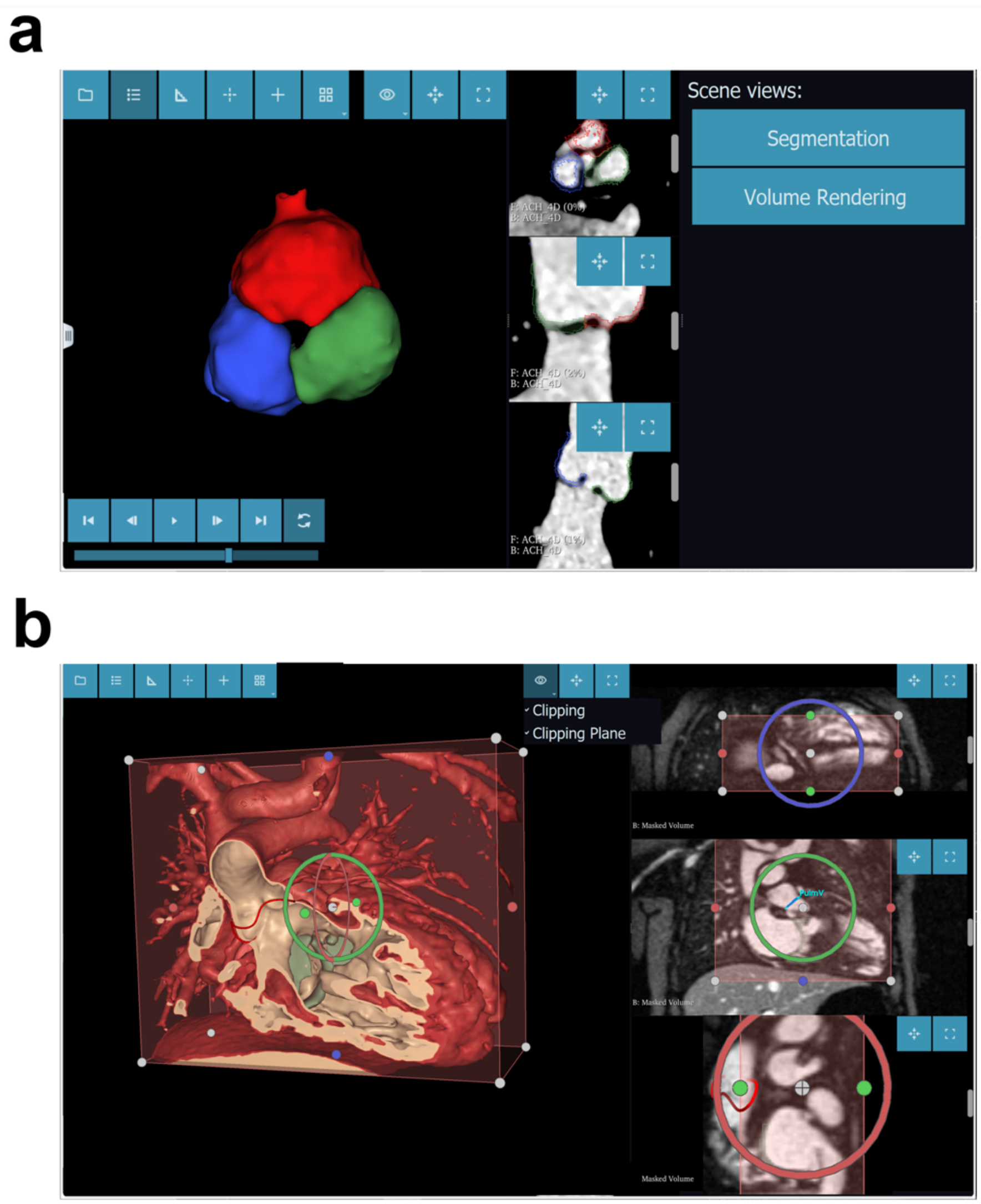

**Fig. 3** SlicerHeart Viewer interface displaying patient-specific cardiac visualization and interactive analysis. (a) Left panel displays CT-derived 3D segmented aortic valve model; central panel shows synchronized multiplanar reformats with real-time correspondence; right panel displays the 'Scene Views' feature enabling toggling between pre-defined models. (b) Left panel shows an interactive 3D cropping box (red) overlaid on myocardial tissue volume rendering, allowing user-defined ROI extraction from all orthogonal planes; right panel displays the corresponding ROI boundaries projected onto synchronized 2D multiplanar reformats, demonstrating real-time 3D-to-2D coordinate synchronization. 2D=2-dimensional; 3D=3-dimensional; CT=computed tomography; ROI=region of interest

a

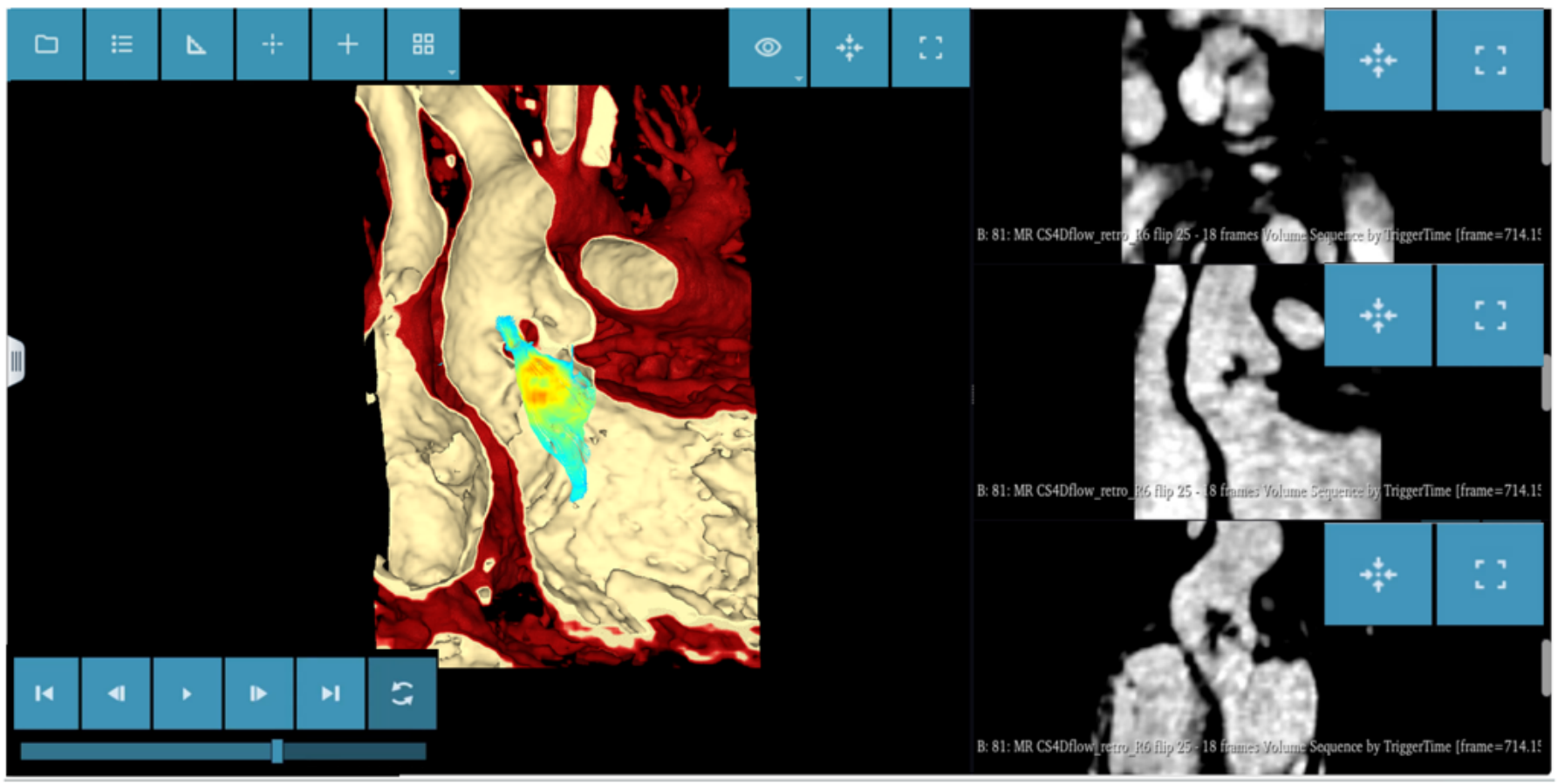

b

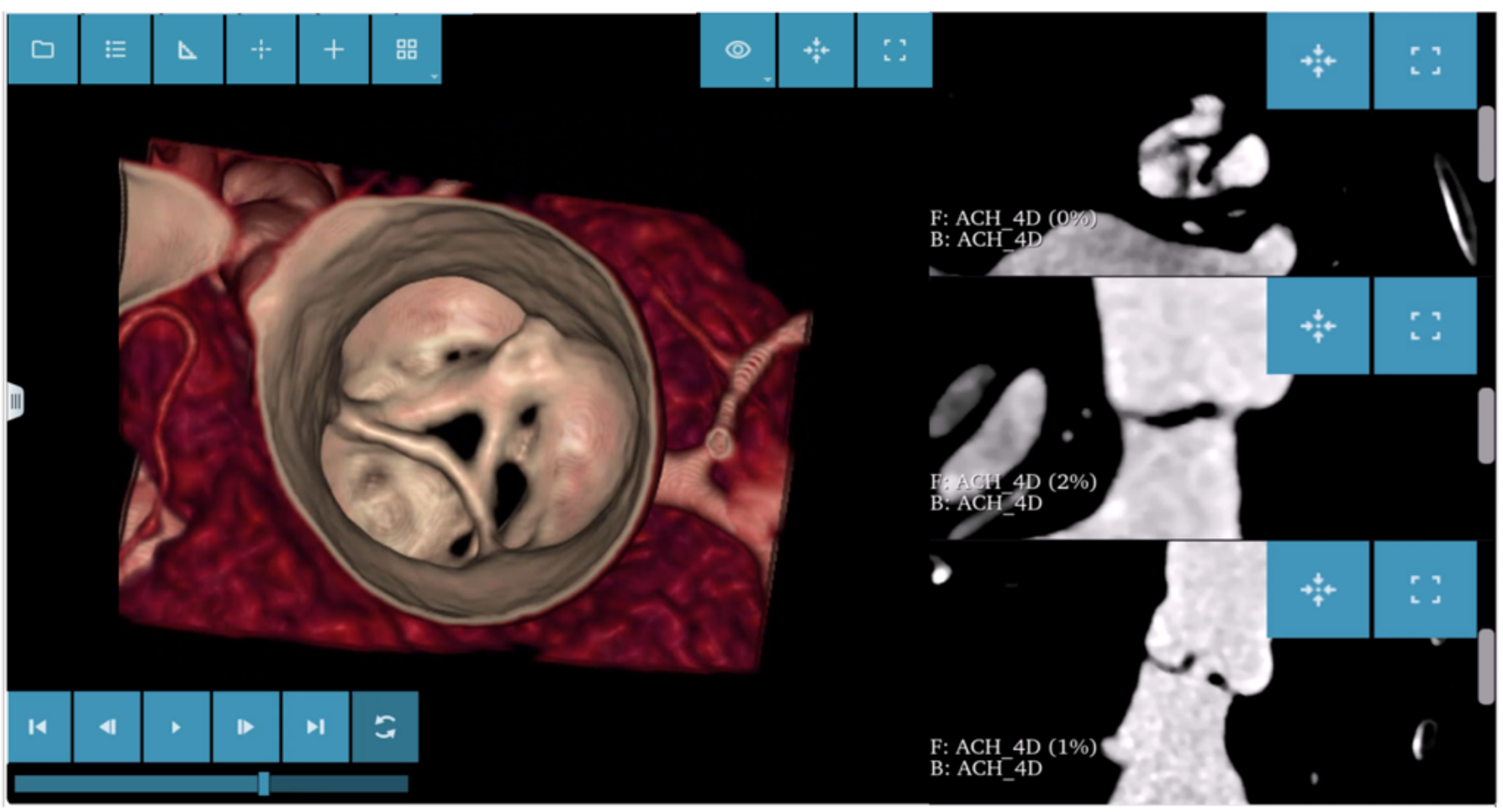

**Fig. 4** SlicerHeart Viewer capability for interactive 4D/5D image visualization and multi-modality data integration. (a) Left panel displays integrated rendering of myocardial tissue with flow streamlines overlaid from 4D flow cardiac MRI, enabling interactive assessment of truncal valve regurgitation throughout the cardiac cycle; right panel shows synchronized multiplanar 2D reformats updated during playback. (b) Left panel displays 4D cardiac CT with tissue rendering for aortic valve coaptation defect assessment; right panel shows synchronized multiplanar 2D reformats updated in real time during playback. 2D=2-dimensional; 4D=4-dimensional; 5D=5-dimensional; CT=computed tomography; MRI=magnetic resonance imaging

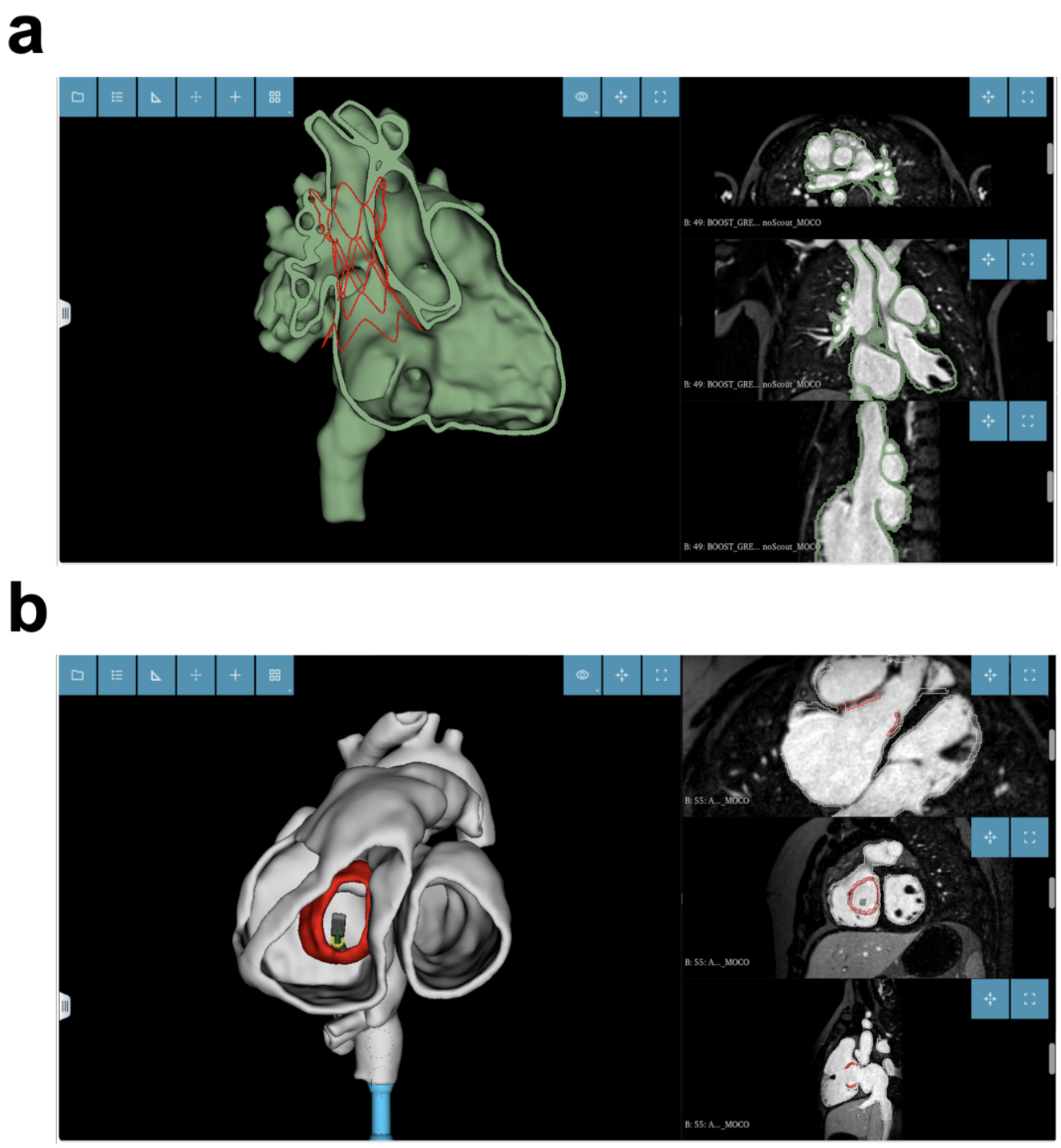

**Fig. 5** SlicerHeart Viewer supporting procedural planning workflows. (a) Left panel displays virtual TPV22 stent positioned within cardiac MRI-derived SVASD/PAPVR heart model; right panel shows synchronized multiplanar 2D reformats. (b) Left panel displays virtual TriClip device in segmented heart and tricuspid valve models to assess feasibility for TEER; right panel shows synchronized multiplanar 2D reformats. 2D=2-dimensional; MRI=magnetic resonance imaging; PAPVR=partial anomalous pulmonary venous return; SVASD=sinus venosus atrial septal defect; TEER=transcatheter edge-to-edge repair

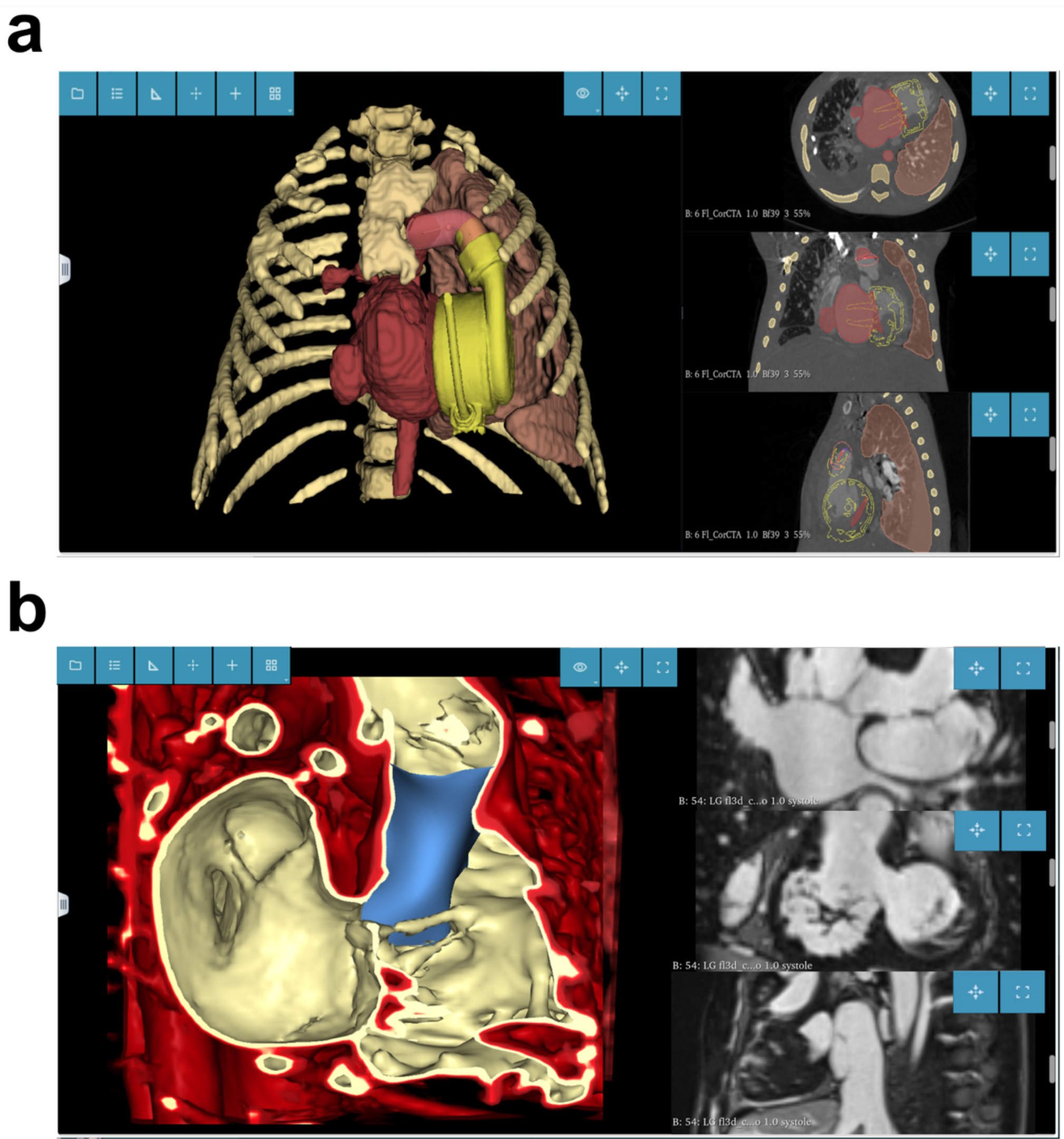

**Fig. 6** SlicerHeart Viewer supporting review of virtually placed devices for surgical planning. (a) Left panel displays virtual HeartMate 3 placed in CT-derived model of rib cage, left lung, and heart; right panel shows synchronized multiplanar 2D reformats. (b) Left panel displays visualization of virtual baffle (blue) placed in cardiac MRI-derived tissue rendering to inform planning of biventricular repair; right panel displays synchronized multiplanar 2D reformats. 2D=2-dimensional; CT=computed tomography; MRI=magnetic resonance imaging

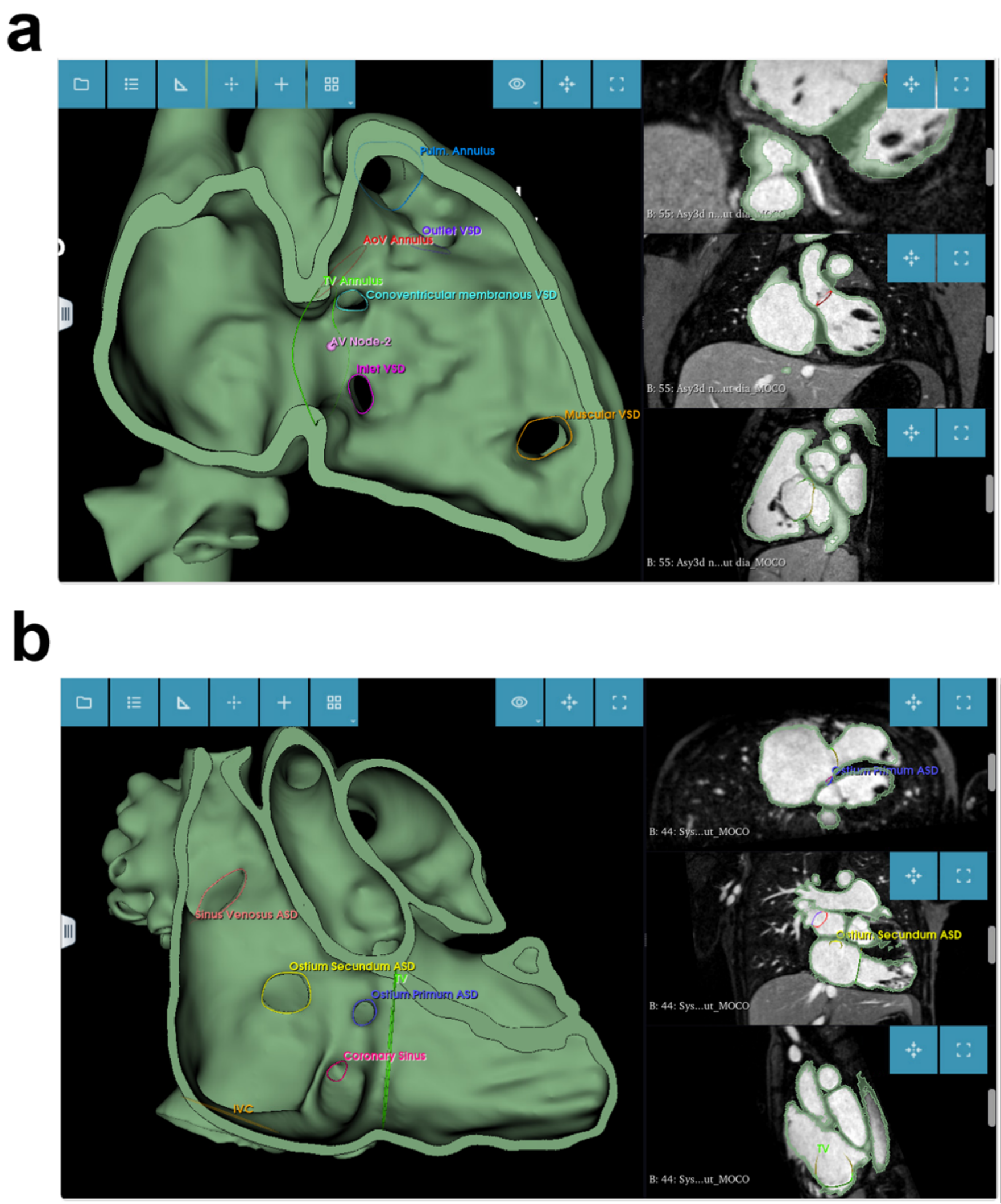

**Fig. 7** SlicerHeart Viewer supporting education and demonstration of congenital heart disease anatomy. (a) Left panel displays a cardiac MRI-derived 3D segmented model with VSD types labeled and highlighted; right panel displays synchronized multiplanar 2D reformats, enabling detailed examination of septal defect morphology. (b) Left panel displays a cardiac MRI-derived 3D segmented model with ASD types labeled and highlighted; right panel displays synchronized multiplanar 2D reformats, demonstrating the relationship between 3D anatomy and cross-sectional imaging planes. 2D=2-dimensional; 3D=3-dimensional; ASD=atrial septal defect; MRI=magnetic resonance imaging; VSD=ventricular septal defect

## Table

| Feature | SlicerHeart Web Viewer | Desktop 3D Slicer SlicerHeart | Commercial Viewer* | Generic web mesh viewer |
|---|---|---|---|---|
| Runs in a web browser (no client install) | Yes | No | Yes | Yes |
| Cross-platform (PC / tablet / phone) | Yes | PC only | Yes | Yes |
| Interactive 3D (rotate, pan, zoom) | Yes | Yes | Yes | Yes |
| Volume rendering | Yes | Yes | No | No |
| Time-resolved 4D / 5D data | Yes | Yes | No | No |
| Capped cross-sections when cropping through segmentations | Yes | Yes | No | No |
| Pre-configured, one-click scenes (saved views, crops, visualizations) | Yes | Yes | Varies | No |
| Interactive cropping and simple measurement by end user | Yes | Yes | Limited | Limited |
| Custom / extensible workflows (e.g., baffle planner, TEER, device modeling) | Yes | Yes | No | No |
| Access to full underlying platform (access to full Slicer) | Yes | Yes | No | No |
| Self-hosted / institutionally hosted (models remain within institutional security infrastructure) | Yes | Yes | No | Varies |
| End-user / institutional licensing cost | Free / Open Source | Free / Open Source | Commercial license | Varies |

**Table 1** Comparison of the SlicerHeart Viewer with existing approaches to reviewing patient-specific cardiac models. *Capabilities of commercial viewers may vary and evolve.

## Online Resource Captions

**Online Resource 1** Video overview of representative clinical cases in the web-based Viewer, including procedural planning models and time-resolved CT, MRI, and 4D flow visualizations. 4D=4-dimensional; CT=computed tomography; MRI=magnetic resonance imaging

**Online Resource 2** Video demonstration of the web-based Viewer on a tablet, showing touch and stylus interaction with representative procedural planning models and time-resolved imaging visualizations.

## References

1. Hoffman JI, Kaplan S: The incidence of congenital heart disease. J Am Coll Cardiol 39:1890-1900, 2002

2. Triedman JK, Newburger JW: Trends in Congenital Heart Disease: The Next Decade. Circulation 133:2716-2733, 2016

3. Lasso A, Herz C, Nam H, Cianciulli A, Pieper S, Drouin S, Pinter C, St-Onge S, Vigil C, Ching S, Sunderland K, Fichtinger G, Kikinis R, Jolley MA: SlicerHeart: An open-source computing platform for cardiac image analysis and modeling. Front Cardiovasc Med 9:886549, 2022

4. Anwar S, Singh GK, Miller J, Sharma M, Manning P, Billadello JJ, Eghtesady P, Woodard PK: 3D Printing is a Transformative Technology in Congenital Heart Disease. JACC Basic Transl Sci 3:294-312, 2018

5. Materialise. Mimics. Available at https://www.materialise.com/en/healthcare/mimics. Accessed 21 July 2026.

6. Fedorov A, Beichel R, Kalpathy-Cramer J, Finet J, Fillion-Robin JC, Pujol S, Bauer C, Jennings D, Fennessy F, Sonka M, Buatti J, Aylward S, Miller JV, Pieper S, Kikinis R: 3D Slicer as an image computing platform for the Quantitative Imaging Network. Magn Reson Imaging 30:1323-1341, 2012

7. Yushkevich PA, Piven J, Hazlett HC, Smith RG, Ho S, Gee JC, Gerig G: User-guided 3D active contour segmentation of anatomical structures: significantly improved efficiency and reliability. Neuroimage 31:1116-1128, 2006

8. Nolden M, Zelzer S, Seitel A, Wald D, Muller M, Franz AM, Maleike D, Fangerau M, Baumhauer M, Maier-Hein L, Maier-Hein KH, Meinzer HP, Wolf I: The Medical Imaging Interaction Toolkit: challenges and advances: 10 years of open-source development. Int J Comput Assist Radiol Surg 8:607-620, 2013

9. Updegrove A, Wilson NM, Merkow J, Lan H, Marsden AL, Shadden SC: SimVascular: An Open Source Pipeline for Cardiovascular Simulation. Ann Biomed Eng 45:525-541, 2017

10. Vigil C, Lasso A, Ghosh RM, Pinter C, Cianciulli A, Nam HH, Abid A, Herz C, Mascio CE, Chen J, Fuller S, Whitehead K, Jolley MA: Modeling Tool for Rapid Virtual Planning of the Intracardiac Baffle in Double-Outlet Right Ventricle. Ann Thorac Surg 111:2078-2083, 2021

11. Iacovella J, Vaiyani D, Pressley S, Lasso A, Sulentic AM, Cianciulli AR, Koluda A, Daemer M, Fogel MA, Whitehead KK, Jolley MA: Rapid Visualization of Valves and Myocardium Using Volume Rendering of 3D Cardiac MRI, 4D Cine, and 4D Flow Images. Radiol Cardiothorac Imaging 8:e250129, 2026

12. 3D Slicer Community Forum. New feature: Support of 5D images and transforms. Available at https://discourse.slicer.org/t/new-feature-support-of-5d-images-and-transforms/45588. Accessed 9 October 2026.
13. McCormick M, Liu X, Jomier J, Marion C, Ibanez L: ITK: enabling reproducible research and open science. Front Neuroinform 8:13, 2014
14. Jolley MA, Sulentic A, Amin S, Gupta M, Ching S, Cianciulli A, Wang Y, Sabin P, Zelonis C, Daemer M, Silvestro E, Coleman K, Ford LK, Edelson JB, Ruckdeschel ES, Cohen MS, Nicolson SC, Gillespie MJ: Introduction of transcatheter edge-to-edge repair in patients with congenital heart disease at a children's hospital. Catheter Cardiovasc Interv 103:326-334, 2024
15. Nam HH, Herz C, Lasso A, Drouin S, Posada A, Morray B, O'Byrne ML, Paniagua B, Joffe D, Mackensen B, Rogers L, Fichtinger G, Jolley MA: Simulation of transcatheter atrial and ventricular septal defect device closure within three-dimensional echocardiography-derived heart models on screen and in virtual reality. J Am Soc Echocardiogr 33:641-644.e2, 2020
16. Omnissa. Horizon 8. Available at https://www.omnissa.com/products/horizon-8/. Accessed 21 July 2026.
17. Kitware Inc. vtk.js. Available at https://kitware.github.io/vtk-js/. Accessed 14 October 2026.
18. Jourdain S, O'Leary P, Schroeder W, Polys NF: Trame: Platform Ubiquitous, Scalable Integration Framework for Visual Analytics. IEEE Comput Graph Appl 45:126-134, 2025
19. Harris PA, Taylor R, Minor BL, Elliott V, Fernandez M, O'Neal L, McLeod L, Delacqua G, Delacqua F, Kirby J, Duda SN, REDCap Consortium: The REDCap consortium: Building an international community of software platform partners. J Biomed Inform 95:103208, 2019
20. Harris PA, Taylor R, Thielke R, Payne J, Gonzalez N, Conde JG: Research electronic data capture (REDCap)—a metadata-driven methodology and workflow process for providing translational research informatics support. J Biomed Inform 42:377-381, 2009
21. Ghosh RM, Jolley MA, Mascio CE, Chen JM, Fuller S, Rome JJ, Silvestro E, Whitehead KK: Clinical 3D modeling to guide pediatric cardiothoracic surgery and intervention using 3D printed anatomic models, computer aided design and virtual reality. 3D Print Med 8:11, 2022